\documentclass[aps,superscriptaddress, nofootinbib, showpacs,12pt]{revtex4}
\usepackage{amsmath,amssymb,amsfonts, bm, natbib}
\usepackage{dsfont}
\usepackage{epsfig}
\usepackage{graphicx}
\usepackage{slashed}
\usepackage{color}

\usepackage{caption}
\usepackage{subcaption}
\usepackage{slashed} 
\usepackage{mathrsfs}

\usepackage{commath}
\usepackage{calc}

\newcommand{\be}{\begin{equation}}
\newcommand{\ee}{\end{equation}}

\newcommand{\bea}{\begin{eqnarray}}
\newcommand{\eea}{\end{eqnarray}}

\newcommand{\bfig}{\begin{figure}}
\newcommand{\efig}{\end{figure}}

\makeatletter
\newcommand*{\rom}[1]{\expandafter\@slowromancap\romannumeral #1@}
\makeatother

\begin{document}
%\preprint{}
\title{Precise predictions for Dirac neutrino mixing}
\author{Gauhar Abbas}
\email{gauhar@prl.res.in}
\affiliation{IFIC, Universitat de Val\`encia -- CSIC, Apt. Correus 22085, 
E-46071 Val\`encia, Spain}
\author{Mehran Zahiri Abyaneh}
\email{Mehran.Za@ific.uv.es}
\affiliation{IFIC, Universitat de Val\`encia -- CSIC, Apt. Correus 22085, 
E-46071 Val\`encia, Spain}
\author{Rahul Srivastava}
\email{Rahul.Srivastava@ific.uv.es}
\affiliation{IFIC, Universitat de Val\`encia -- CSIC, Apt. Correus 22085, 
E-46071 Val\`encia, Spain}

%%%%%%%%%%%%%%%%%%%%%%%%%%%%%%%%%%%%%%%%%%%%%%%%%%%%%%%%%%%%%%%%%%%%%%%%%%%%%%%%%%%%%%%%%%%%%%%%%%%%%%%%%%%%%%%%%%%%%%%%%%%%%%%%%%%%%%%%%%%%%%%%%%%%%%%%%%%%%%%%%%%%%%%%%%%%%%%%%%%%%%%%%%%%%%%%%%%%%%%%%%%%%%%%%%%%%%%%%%%%%%%%%%%%%%%%%%%%%%%%%%%%%%%%%%%%%%%%%%%%%%

\begin{abstract}
The neutrino mixing parameters are thoroughly studied using  renormalization-group evolution of Dirac neutrinos with recently proposed parametrization of the neutrino mixing angles referred as `high-scale mixing relations'. The correlations among all neutrino mixing and $CP$ violating observables are investigated.  The predictions for the neutrino mixing angle $\theta_{23}$ are precise, and could be easily tested by ongoing and future experiments.  We observe that the high scale mixing unification hypothesis is incompatible with Dirac neutrinos due to updated experimental data. 
\end{abstract}

\pacs{14.60.Pq, 11.10.Hi, 11.30.Hv, 12.15.Lk}

\maketitle

%%%%%%%%%%%%%%%%%%%%%%%%%%%%%%%%%%%%%%%%%%%%%%%%%%%%%%%%%%%%%%%%%%%%%%%%%%%%%%%%%%%%%%%%%%%%%%%%%%%%%%%%%%%%%%%%%%%%%%%%%%%%%%%%%%%%%%%%%%%%%%%%%%%%%%%%%%%%%%%%%%%%%%%%%%%%%%%%%%%%%%%%%%%%%%%%%%%%%%%%%%%%%%%%%%%%%%%%%%%%%%%%%%%%%%%%%%%%%%%%%%%%%%%%%%%%%%%%%%%%%%
\section{Introduction}
Neutrino mixing is one of the most fascinating and challenging discoveries.  This is starkly different from  quark mixing which is small in the standard model (SM).  There are a number of ways to explain these two very different phenomena.  The quark-lepton unification, which is one of the main attractive features of the grand unified theories (GUT)\cite{Pati:1974yy,Georgi:1974sy,Fritzsch:1974nn}, could provide an explanation of the origin of neutrino and quark mixing since quarks and leptons live in a joint represenation of the symmetry group.  Another interesting approach is to use flavor symmetries \cite{Altarelli:2010gt,King:2013eh,Holthausen:2013vba,Araki:2013rkf,Ishimori:2014jwa}.  These symmetries could also naturally appear in GUT theories\cite{Lam:2014kga}.

To explain the origin of neutrino and quark mixing, recently a new parametrization of the neutrino mixing angles in terms of quark mixing angles was proposed in Ref.\cite{Abbas:2015vba}.  The varoius simplified limits of this prameterization are referred as `high-scale mixing relations'(HSMR).  The parametrization is inspired by the high scale mixing unification (HSMU) hypothesis which states that at certain high scales the neutrino mixing angles are identical to that of the quark mixing angles\cite{Mohapatra:2003tw,Mohapatra:2005gs,Mohapatra:2005pw,Agarwalla:2006dj}. This hypothesis is studied in detail in Refs.\cite{Abbas:2014ala,Abbas:2013uqh,Srivastava:2015tza,Srivastava:2016fhg,Haba:2012ar} .

The HSMR parametrization of the neutrino mixing angles assumes that the neutrino mixing angles are proportional to those of quarks due to some underlying theory which could be a quark-lepton unification or models based on flavor symmetries.  In fact, such models are also presented in Ref.\cite{Abbas:2015vba}.  The scale where the HSMR parametrization could be realized is referred as unification scale. In its most general form, the HSMR parametrization can be written as follows:
\begin{equation}
\label{hsmr}
\theta_{12} = \alpha_1^{k_1} ~\theta_{12}^q, ~~ \theta_{13} = \alpha_2^{k_2}~ \theta_{13}^q, ~~\theta_{23} =\alpha_3^{k_3}  \theta_{23}^q. 
\end{equation}
where $\theta_{ij}$ (with $i,j=1,2,3$) denotes leptonic mixing angles and $\theta_{ij}^q$ 
are the quark mixing angles. Exponents  $k_i$ with $i=(1,2,3)$ are real. Predictions of the HSMR parametrization could be a strong hint of the quark-lepton unification, some flavor symmetry or both.

The HSMR parametrization is studied in the framework of the SM extended by  the minimum supersymmetric
standard model (MSSM).  The beginning point is to run the quark mixing angles from the low scale (mass of the $Z$ boson) to  the 
supersymmetry (SUSY) breaking scale using the renormalization-group (RG) evolution of the SM.  The RG equations of the MSSM govern the evolution of quark mixing angles from the SUSY breaking scale to the unification scale.  After obtaining quark mixing angles at the unification scale, the HSMR parametrization is used to run neutrino mixing parameters from the unification scale to the SUSY breaking scale via RG evolution of the  MSSM. From the SUSY breaking scale to the low scale, the SM RG equations are used to evolve the neutrino mixing parameters.  The free parameters controlling the top-down evolution of the neutrino mixing parameters are masses of the three light neutrinos, Dirac CP phase and parameters $\alpha_i$.  Masses  of neutrinos must be quasidegenerate  and normal hierarchical.  Furthermore, the large value of $\tan \beta$ is required\cite{Abbas:2015vba}.

On the other hand, the nature of neutrinos is still unknown.  They could be equally Dirac or Majorana in nature. Hence, from the phenomenological point of view, Dirac neutrinos are as important as Majorana neutrinos.  There are many ongoing important experiments to test the nature of neutrinos\cite{Agostini:2013mzu,Auger:2012ar,Gando:2012zm,Alessandria:2011rc}.   However, for the Dirac mass of neutrinos, the Yukawa couplings  for neutrinos  seem to be unnaturally small.  The elegant way to explain this fine-tuning is see-saw mechanism which assumes that neutrinos are Majorana in nature\cite{Minkowski,GellMann:1980vs,Yanagida:1979as,Glashow:1979nm,Mohapatra:1979ia}. 

The smallness of masses for Dirac neutrinos could be explained in many models using heavy degrees of freedom\cite{Abbas:2016qqc,Ma:2014qra, Mohapatra:1986bd,ArkaniHamed:2000bq,Borzumati:2000mc,Kitano:2002px,Abel:2004tt,Murayama:2004me,Smirnov:2004hs,Mohapatra:2004vr}.  There are also models based on extra dimensions which explain the smallness of Dirac neutrino mass by a small overlapping of zero-mode profiles along extra dimensions\cite{Hung:2004ac,Ko:2005sh,Antusch:2005kf}.  Dirac neutrinos seem to be a natural choice in certain orbifold compactifications of the heterotic string where the standard see-saw mechanism is difficult to realize\cite{Giedt:2005vx}.  Cosmological data do not prefer Majorana or Dirac neutrinos either.  For instance, the  baryon asymmetry of the Universe can also be explained for Dirac neutrinos in various theoretical models\cite{Dick:1999je,Murayama:2002je,Gu:2006dc,Gu:2007mi,Gu:2007mc,
Gu:2007ug,Gu:2012fg}.  

Although the RG evolution of Majorana neutrinos is extensively studied in the literature\cite{Mohapatra:2003tw,Mohapatra:2005gs,Mohapatra:2005pw, Agarwalla:2006dj,Casas:2003kh,Abbas:2014ala,Abbas:2013uqh,Srivastava:2015tza,Casas:1999tg}, less attention is being paid to the RG evolution of Dirac neutrinos.  In fact, as far as we know,  it was shown for the first time in Ref.\cite{Abbas:2013uqh} that RG evolution for Dirac neutrinos can explain the large neutrino mixing assuming the HSMU hypothesis.  However, as we show later, these results are ruled out by new updated data\cite{Forero:2014bxa,Capozzi:2013csa,Gonzalez-Garcia:2014bfa} and due to an improved algorithm used in the package REAP\cite{private}.

It is established that the HSMR parametrization can explain the observed pattern of the neutrino mixing assuming they are Majorana in nature\cite{Abbas:2015vba}. In this paper, we investigate the consequences of the HSMR parametrization using the RG evolution of Dirac neutrinos.

This paper is organized in the following way: In Sec. \ref{sec1}, we present our results on the RG evolution of  the neutrino mixing parameters. In Sec. \ref{sec2} we present a model with naturally small Dirac neutrino masses, where the HSMR parametrization discussed in Eq.\ref{hsmr} can be explicitly realized.  We summarize our work in Sec. \ref{sec3}.

\section{RG evolution of the neutrino mixing parameters for Dirac neutrinos}
\label{sec1}
Now we present our results. The RG equations describing the evolution of the neutrino mixing parameters for Dirac neutrinos are derived in Ref. \cite{Lindner:2005as}. We have used  Mathematica- based package REAP for the computation of the RG evolution at two loops \cite{Antusch:2005gp}.  The first step is to evolve quark mixing angles, gauge couplings, Yukawa couplings of quarks, and  charged leptons from the low scale to the SUSY breaking scale.  From the SUSY breaking scale to the unification scale, evolution undergoes the MSSM RG equations.  The quark mixing angles at the unification scale after evolution are   $\theta_{12}^q = 13.02^\circ$, $\theta_{13}^q=0.17^\circ$ and $\theta_{23}^q=2.03^\circ$.   Now, quark-mixing angles are used by the HSMR parametrization at the unification scale and neutrino mixing parameters are evolved down to the SUSY breaking scale using the MSSM RG equations.  After this, the evolution of mixing parameters are governed by the SM RG equation.  The value of $\tan \beta$ is chosen to be $55$.  For simplification, we have assumed $k_1=k_2=k_3=1$ in the HSMR parametrization. The global status of the neutrino mixing parameters is given in Table \ref{tab1}.
\begin{table}[h]
\begin{center}
\begin{tabular}{|c|c|c|}
  \hline
  Quantity                                       & Best fit                                   &  3$\sigma$ range\\
  \hline
  $\Delta m^2_{21}~(10^{-5}~{\rm eV}^2)$         & $7.60$                                    &  7.11 -- 8.18       \\
  $\Delta m^2_{31}~(10^{-3}~{\rm eV}^2)$         & $2.48$                                 &  2.30 -- 2.65        \\
  $\theta_{12}^{\circ}$                        & $34.6$                                   &  31.8 -- 37.8           \\
  $\theta_{23}^{\circ}$                        & $48.9$            &  38.8 -- 53.3   \\
  $\theta_{13}^{\circ}$                        & $8.6 $                                  &  7.9-- 9.3    \\
  \hline
     \end{tabular}
\end{center}
\caption{The global fits for the neutrino mixing parameters \cite{Forero:2014bxa}}
 \label{tab1}
\end{table}

\subsection{Results for the SUSY breaking scale at 2 TeV}
In this subsection, we present our results for the SUSY breaking scale at 2 TeV following the direct LHC searches \cite{Craig:2013cxa}.   The unification scale where the HSMR parametrization could be realized is chosen to be GUT scale ($2 \times 10^{16}$ GeV).  The free parameters of the analysis are shown in Table \ref{tab2}. 

\begin{table}[h]
\begin{center}
\begin{tabular}{|c|c|}
  \hline
  Quantity                                       & Range at the unification scale                               \\
  \hline
  $\alpha_1$         & $0.7 - 0.8$                                       \\
  $\alpha_2$         & $2.12 - 2.78$                                    \\
  $\alpha_3$                        & $1.002 -1.01$                                           \\
  $m_1$(eV)                        & $0.49227 - 0.49825$            \\
  $m_2$ (eV)                       & $0.494 - 0.5$                                  \\
   $m_3$ (eV)                       & $0.52898 - 0.53542$                                   \\
    $\delta_{Dirac}$                        & $(-14^\circ,  14^\circ)$                                   \\
  \hline
     \end{tabular}
\end{center}
\caption{The free parameters of the analysis chosen at the unification scale.}
 \label{tab2}
\end{table}

\begin{figure}[htb]
\centering
\includegraphics[width=6.5cm, height=5cm]{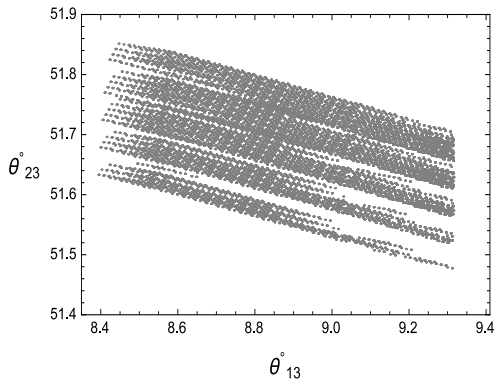}
\caption{ The variation of $\theta_{23}^\circ$ with respect to $\theta_{13}^\circ$.}
\label{fig1}
\end{figure}

In Fig. \ref{fig1}, we show a  correlation between mixing angles $\theta_{13}$ and $\theta_{23}$. It is obvious that our prediction for  $\theta_{23}$ is  precise.  The allowed range of $\theta_{13}$ is $7.94^\circ - 9.3^\circ$.  The corresponding range of $\theta_{23}$ is $51.5^\circ - 52.64^\circ$. It is important to note that the predictions for  $\theta_{13}$  include the best fit value.   Another important prediction is that  $\theta_{23}$ is nonmaximal and lies in the second octant.  Being precise, this correlation is easily testable in future and ongoing experiments such as INO, T2K, NO$\nu$A, LBNE, 
Hyper-K, and PINGU \cite{Abe:2011ks,Patterson:2012zs,Adams:2013qkq,Ge:2013ffa,Kearns:2013lea,Athar:2006yb}.

\begin{figure}[htb]
\begin{minipage}[b]{0.45\linewidth}
\vspace*{0.65cm}
\centering
\includegraphics[width=6.5cm, height=5cm]{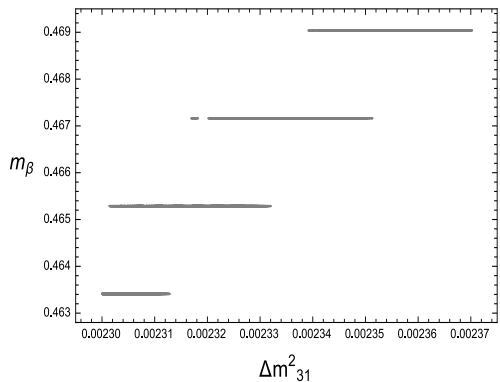}
\caption{ The variation of $m_\beta$   with respect to $\Delta m_{31}^2$.}
\label{fig2}
\end{minipage}
\end{figure}

In Fig. \ref{fig2}, we show the variation of  ``averaged electron neutrino mass'' $m_\beta$ \cite{Drexlin:2013lha} with respect to $\Delta m_{31}^2$.  The allowed range of $m_\beta$ is $0.4633-0.4690$ eV which is precise.  The upper bound on  $m_\beta$ is  $2$ eV from tritium beta decay \cite{Kraus:2004zw,Aseev:2011dq}.  The KATRIN experiment  is expected to probe $m_\beta$ as low as $0.2$ eV at $90\%$ C.L. \cite{Drexlin:2013lha}. Hence, our prediction for $m_\beta$ is well within the reach of the KATRIN experiment.  The allowed range for $\Delta m_{31}^2$ is $(2.30 - 2.37) \times 10^{-3} \textrm{eV}^2$ which is bounded with respect to the $3\sigma$ range given by the global fit in Table \ref{tab1}.  It should be noted that the best fit value of $\Delta m_{31}^2$ given in Table \ref{tab1} is excluded by our results.

\begin{figure}[htb]
\centering
\includegraphics[width=6.5cm, height=5cm]{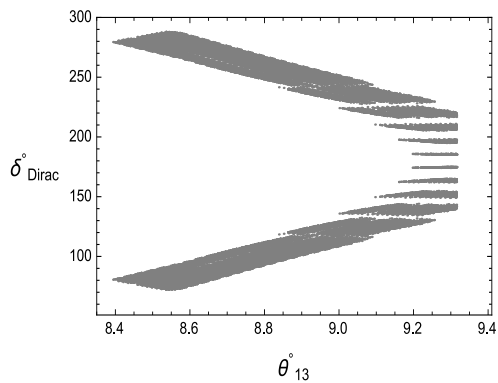}
\caption{ The variation of $\delta_{Dirac}^\circ$ with respect to $\theta_{13}^\circ$.}
\label{fig3}
\end{figure}

\begin{figure}[htb]
\centering
\includegraphics[width=6.5cm, height=5cm]{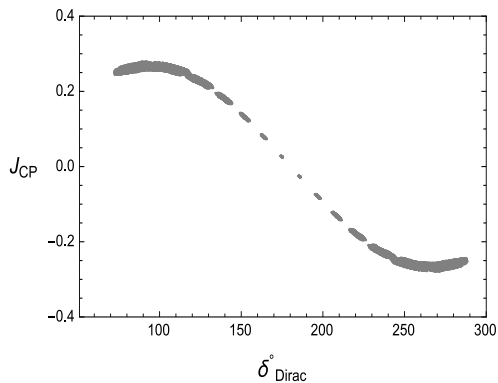}
\caption{ The variation of $J_{CP}$ with respect to $\delta_{Dirac}^\circ$.}
\label{fig4}
\end{figure}

We show in Fig.\ref{fig3}  another important predictions of this work.   This is the variation of the $CP$ violating Dirac phase $\delta_{Dirac}$ with respect to  $\theta_{13}$.  The Dirac phase $\delta_{Dirac}$ is not known from experiments.  Hence, any prediction of this important observable is of great interest.  Our prediction for  $\delta_{Dirac}$ is $80.01^\circ ~\textrm{to} ~287.09^\circ$ excluding a sufficient part of the allowed parameter space of this quantity.  In Fig.\ref{fig4}, we show the behavior of the Jarlskog invariant $J_{CP}$ with respect to  Dirac phase $\delta_{Dirac}$.  The allowed range for this observable is $-0.266~\textrm{to}  ~0.266$.  Thus, a large $CP$ violation is possible in our analysis.

\begin{figure}[htb]
\centering
\includegraphics[width=6.5cm, height=5cm]{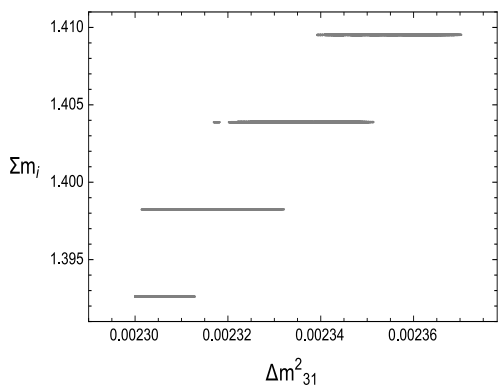}
\caption{ The variation of $\Sigma m_i$ with respect to $\Delta m_{31}^2$.}
\label{fig5}
\end{figure}

The variation of the sum of three neutrino masses, $\Sigma m_i$ with respect to $\Delta m_{31}^2$ is shown in Fig.\ref{fig5}. The allowed range of $\Sigma m_i$ is $1.393-1.410$ eV, which is precise.  We comment that our prediction for $\Sigma m_i$  is a little higher than that provided by the cosmological and astrophysical observations which is $0.72$ eV at $95\% $C.L. \cite{Ade:2015xua}. However,  cosmological limit on  $\Sigma m_i$ is highly model dependent.  For example, as shown in Fig. 29  of Ref.\cite{Ade:2015xua} this could be as large as $1.6$eV. Furthermore, Ref.\cite{Ade:2015xua} assumes degenerate neutrinos ignoring the observed mass splittings whereas their model ($\Lambda$CDM) assumes two massless and one massive neutrino with $\Sigma m_i =0.06$eV.  Moreover, $\Lambda$CDM is facing several challenges in explaining structures on galaxy scales\cite{Famaey:2013ty}. Hence, our predictions are aimed to test in laboratory-based experiments like KATRIN\cite{Drexlin:2013lha}.

We do not obtain any constraints on the mixing angle $\theta_{12}$ and mass square difference $\Delta m_{21}^2$.  The whole $3\-\sigma$ ranges of global fit are allowed in this case for these quantities.  

\subsection{Variation of the SUSY breaking scale}
Now, we discuss the effect of variation of the SUSY breaking scale on our predictions.  In this case, we change the SUSY breaking scale to 5 TeV.  However, the unification scale is still at the GUT scale. Our results are summarized in Tables \ref{tab3} and \ref{tab4}.  In Table \ref{tab3}, we provide our free parameters which are chosen at  the GUT scale.  Our predictions at the low scale are given in Table \ref{tab4}.

\begin{table}[h]
\begin{minipage}[b]{0.45\linewidth}
\vspace*{0.65cm}
\begin{center}
\begin{tabular}{|c|c|}
  \hline
  Quantity                                       & Range at the unification scale                               \\
  \hline
  $\alpha_1$         & $0.88 - 1.012$                                       \\
  $\alpha_2$         & $2.72 - 2.85$                                    \\
  $\alpha_3$                        & $1.095$                                           \\
  $m_1$(eV)                        & $0.46878 - 0.47380$            \\
  $m_2$ (eV)                       & $0.47 - 0.475$                                  \\
   $m_3$ (eV)                       & $0.50321 - 0.50857$                                   \\
    $\delta_{Dirac}$                        & $(-14^\circ,14^\circ)$                                   \\
  \hline
     \end{tabular}
\end{center}
\caption{The free parameters of the analysis chosen at the unification scale for the SUSY breaking scale at 5 TeV.}
 \label{tab3}
\end{minipage}
\hspace{0.7cm}
\begin{minipage}[b]{0.45\linewidth}
\begin{center}
\begin{tabular}{|c|c|}
  \hline
  Quantity                                       & Range at the low scale                                \\
  \hline
  $\theta_{12}$         & $32.85^\circ - 37.74^\circ$                                       \\
  $\theta_{13}$         & $7.94^\circ - 8.20^\circ$                                    \\
  $\theta_{23}$                        & $38.86^\circ -39.45^\circ$                                           \\
  $m_1$(eV)                        & $0.44458 - 0.44932$            \\
 $\Delta m^2_{21}~(10^{-5}~{\rm eV}^2)$                        &    $7.15 - 8.15$                              \\
  $\Delta m^2_{31}~(10^{-3}~{\rm eV}^2)$                      &$2.30- 2.34 $                                    \\
    $m_\beta$ (eV)                       & $0.4447 - 0.4468 $                                    \\
    $\Sigma m_i$ (eV)                &      $1.337 - 1.351$                                                 \\
      $\delta_{Dirac}$                        & $281.28^\circ - 355.49^\circ$  {\rm and} $0 - 89.14^\circ$                                \\
       $J_{CP}$                        & $-0.2511~ {\rm to} ~0.2511$                                \\
  \hline
     \end{tabular}
\end{center}
\caption{Predictions of neutrino mixing parameters and other observables at the low scale for the SUSY breaking scale at 5 TeV.}
 \label{tab4}
\end{minipage}
\end{table}  

We observe that  the mixing angle $\theta_{12}$ and mass square difference  $\Delta m^2_{21}$ were unconstrained for the SUSY breaking scale at 2 TeV in the previous subsection.  Now, we observe that these quantities are bounded with respect to the $3\sigma$ range given by the global  fit. The mixing angle $\theta_{23}$, unlike the investigation for SUSY breaking scale 2 TeV,  lies in the first octant and is non-maximal.  

\subsection{Variation of the unification scale}
In this subsection, we investigate the variation of the unification scale. In Tables \ref{tab5} and \ref{tab6}, we show our results when we choose the unification scale to be $10^{12}$ GeV which is well below the GUT scale.  However, the SUSY breaking scale is kept to 2 TeV.  We show in Table \ref{tab5}, the values of the free parameters chosen at the unification scale.  In Table \ref{tab6}, we present our results.  The first remarkable prediction is the sum of neutrino masses which is well below the cosmological bound.  The Dirac $CP$  phase has a precise range.  The mixing angle $\theta_{12}$ and mass square difference  $\Delta m^2_{21}$ are now relatively constrained.  The mixing angle $\theta_{23}$ lies in the first octant, and is nonmaximal.

\begin{table}[h]
\begin{minipage}[b]{0.45\linewidth}
\vspace*{0.65cm}
\begin{center}
\begin{tabular}{|c|c|}
  \hline
  Quantity                                       & Range at the unification scale                               \\
  \hline
  $\alpha_1$         & $0.67 - 0.85$                                       \\
  $\alpha_2$         & $19.9 - 20.92$                                    \\
  $\alpha_3$                        & $7.41 - 7.42$                                           \\
  $m_1$(eV)                        & $0.19815 - 0.20311$            \\
  $m_2$ (eV)                       & $0.2 - 0.205 $                                  \\
   $m_3$ (eV)                       & $0.21100 - 0.21628 $                                   \\
    $\delta_{Dirac}$                        & $(-10^\circ,  18^\circ)$                                   \\
  \hline
     \end{tabular}
\end{center}
\caption{The free parameters of the analysis chosen at the unification scale  of  $10^{12}$ GeV and SUSY breaking scale of 2 TeV.}
 \label{tab5}
\end{minipage}
\hspace{0.7cm}
\begin{minipage}[b]{0.45\linewidth}
\begin{center}
\begin{tabular}{|c|c|}
  \hline
  Quantity                                       & Range at the low scale                                \\
  \hline
  $\theta_{12}$         & $32.35^\circ - 37.34^\circ$                                       \\
  $\theta_{13}$         & $7.94^\circ - 8.45^\circ$                                    \\
  $\theta_{23}$                        & $38.83^\circ -39.18^\circ$                                           \\
  $m_1$(eV)                        & $0.18321 - 0.18801$            \\
 $\Delta m^2_{21}~(10^{-5}~{\rm eV}^2)$                        &    $7.77 - 8.17$                              \\
  $\Delta m^2_{31}~(10^{-3}~{\rm eV}^2)$                      &$2.30- 2.42 $                                    \\
    $m_\beta$ (eV)                       & $0.1834 - 0.1880 $                                    \\
    $\Sigma m_i$ (eV)                &      $0.556 - 0.570$                                                 \\
      $\delta_{Dirac}$                        & $182.66^\circ - 203.43^\circ$ {\rm and} $0-120^\circ$                                 \\
       $J_{CP}$                        & $-0.1020~ {\rm to} ~0.2336$                                   \\
  \hline
     \end{tabular}
\end{center}
\caption{Predictions of neutrino mixing parameters and other observables for the unification scale  of  $10^{12}$ GeV  and the SUSY breaking scale at 2 TeV.}
 \label{tab6}
\end{minipage}
\end{table}  

We conclude that there is no parameter space beyond the GUT scale for Dirac neutrinos so that we could recover the experimental data at the low scale using the RG evolution.  This is a strong prediction and could be useful in construction of models (particularly GUT models) where Dirac neutrinos are the natural choice\cite{Ma:2014qra, Mohapatra:1986bd,ArkaniHamed:2000bq,Borzumati:2000mc,Kitano:2002px,Abel:2004tt,Murayama:2004me,Smirnov:2004hs,Mohapatra:2004vr}.

\section{ Model for the HSMR parametrization} 
\label{sec2}
We have investigated the HSMR parametrization for Dirac neutrinos in a model independent way.  However, for the sake of completeness, in this section we discuss theoretical implementation of the HSMR parametrization in a specific model for Dirac neutrinos. Our model is based on a model presented in Ref.  \cite{Haba:2012ar, Haba:2011pm} which provides Dirac neutrinos with naturally small masses. This model is a type of neutrinophilic SUSY extension of the SM which can easily be embedded in a class of $SU(5)$ models.

To obtain HSMR parametrization in the model given  in Ref.  \cite{Haba:2011pm}, we impose a $Z_3$ discrete symmetry on this model.  Under the $Z_3$ symmetry the first generation of both left- and right-handed quarks and leptons transforms as $1$, while the second generation transforms as $\omega$ and the third generation transforms as $\omega^2$, where $\omega$ denotes cube root of unity with $\omega^3 = 1$. All other fields transform trivially as $1$ under the $Z_3$ symmetry. The $Z_3$ symmetry ensures that 
the mass matrices for both up and down quarks as well as for charged leptons and neutrinos are all simultaneously diagonal. This in turn implies that the $V_{\rm{CKM}}$ as well as $V_{\rm{PMNS}}$ are both unity and there is no generation mixing in either quark or lepton sectors.

To allow for the mixing, we break $Z_3$ in a way as done in Ref. \cite{Ma:2002yp}. Such corrections can arise from the soft SUSY breaking sector\cite{Babu:2002dz,Babu:1998tm, Gabbiani:1996hi}.  For this purpose, we allow symmetry breaking terms of the form $|y''_i| <<|y'_i| <<|y_i| $ where $|y_i|$ are the terms invariant under $Z_3$ symmetry, and $|y'_i|, |y''_i|$ are the symmetry breaking terms transforming as $\omega, \omega^2$ under the $Z_3$ symmetry.  This symmetry breaking pattern is well established and is known  to explain the CKM structure of the quark sector\cite{Ma:2002yp}.  Here, we have imposed this pattern on quarks as well as leptons simultaneously.

Including these symmetry breaking terms, the mass matrices for quarks and leptons become 

\begin{eqnarray}
   M_{u,d,l} = \left( 
\begin{array}{ccc}
y_1 v           &  y'_2  v         &   y''_3 v    \\
y''_1 v         &  y_2 v           &   y'_3  v     \\
y'_1 v          &  y''_2 v         &   y_3 v       \\  
\end{array}
\right)~,  \qquad
M_{\nu} = \left( 
\begin{array}{ccc}
y_1 u           &  y'_2  u         &   y''_3 u    \\
y''_1 u         &  y_2 u           &   y'_3  u     \\
y'_1 u          &  y''_2 u         &   y_3 u       \\ 
\end{array}
\right)~,
\label{brok-mass-mat}
\end{eqnarray}
where $v$ stands for the vacuum expectation value (vev) of the usual $H_u, H_d$ doublet scalars of MSSM and $u$ is the vev of the neutrinophilic scalar $H_\nu$ as discussed in Ref. \cite{Haba:2011pm}. Also, for the sake of brevity we have dropped the sub- and superscripts on the various terms. The mass matrix in (\ref{brok-mass-mat})
is exactly same as the mass matrix obtained in Ref. \cite{Ma:2002yp} and can be diagonalized in the same way as done in Ref.\cite{Ma:2002yp}.  The mass matrices of (\ref{brok-mass-mat}) lead to a ``Wolfenstein-like structure'' for both CKM and Pontecorvo-Maki-Nakagawa-Sakata (PMNS) matrices, thus leading to the HSMR parametrization given in Eq.\ref{hsmr}.  Since this model is a modification of model given in Ref.\cite{Haba:2011pm} which can be embedded in a class of $SU(5)$ GUT models, therefore, it can also be easily embedded in the $SU(5)$ GUT model in a quite similar way as done in Ref.\cite{Haba:2011pm}.

\section{Summary}
\label{sec3}
Neutrino mixing is remarkably different from small quark mixing.  The aim of the present work is to provide an insight into a common origin of neutrino as well as quark mixing for Dirac neutrinos.  Furthermore,  we show that smallness of neutrino masses can be explained through the RG evolution of Dirac neutrinos. The HSMR parametrization of neutrino mixing angles is one among many other theoretical frameworks constructed for this purpose.  The origin of this parametrization lies in the underlying concept of the quark-lepton unification or flavor symmetries or both.  Hence, the confirmation of predictions provided by the HSMR parametrization would be a strong hint of the quark-lepton unification or a grand symmetry operating at the unification scale.

As far as our knowledge is concerned, it was shown for the first time in Ref.\cite{Abbas:2013uqh} that the RG evolution can also explain the large neutrino mixing  for Dirac neutrinos.  However, as we have shown in this work, these results are no longer valid due to updated  experimental data\cite{Forero:2014bxa,Capozzi:2013csa,Gonzalez-Garcia:2014bfa} and the improved algorithm used in the package REAP\cite{private}. 

In the present work, we have investigated the RG evolution of Dirac neutrinos in the framework of the HSMR parametrization.  To our knowledge, this is the first thorough study on the RG behavior of Dirac neutrinos.  The main achievement is that the RG evolution of Dirac neutrinos could explain the large neutrino mixing including the observation of a small and nonzero value of the mixing angle $\theta_{13}$. We obtain strong correlations among different experimental observables.  Our predictions for the mixing angles  $\theta_{13}$,  $\theta_{23}$, averaged electron neutrino mass $m_\beta$, Dirac $CP$ phase $\delta_{Dirac}$ and the sum of three neutrino masses, $\Sigma m_i$ are precise and easily testable at ongoing and future experiments like INO, T2K, NO$\nu$A, LBNE, 
Hyper-K, PINGU and KATRIN \cite{Abe:2011ks,Patterson:2012zs,Adams:2013qkq,Ge:2013ffa,Kearns:2013lea,Athar:2006yb,Drexlin:2013lha}.  The mixing angle $\theta_{23}$ is nonmaximal and lies in the second octant for the SUSY breaking scale 2 TeV and unification scale at the GUT scale.  For the variation of the SUSY breaking scale and the unification scale, the mixing angle $\theta_{23}$ is nonmaximal and lies in the first octant. The predictions for the mass square difference  $\Delta m_{31}^2$ are also well constrained and testable in experiments.   Furthermore,  the Dirac $CP$ phase is found to be lying in precise ranges in our analysis.  The unification scale beyond the GUT scale is ruled out in our investigation.  This fact could be useful for the GUT theories having Dirac neutrinos\cite{Ma:2014qra, Mohapatra:1986bd,ArkaniHamed:2000bq,Borzumati:2000mc,Kitano:2002px,Abel:2004tt,Murayama:2004me,Smirnov:2004hs,Mohapatra:2004vr}.  We remark that we have investigated the RG evolution of neutrino mixing parameters at two loops.  This is a crucial input since the RG evolution at one loop is insufficient to provide the required enhancement of the mixing angles which in turn, cannot yield the results obtained in this work.

One of the main consequences of our investigation is that the HSMU hypothesis is not compatible with Dirac neutrinos due to updated experimental data\cite{Forero:2014bxa,Capozzi:2013csa,Gonzalez-Garcia:2014bfa} and a better algorithm used in the package REAP\cite{private}.   The HSMU hypothesis is a  particular realization of the HSMR parametrization when we choose $\alpha_1= \alpha_2 =\alpha_3=1$ for $k_1= k_2 =k_3=1$.   As can be observed from Tables \ref{tab2}, \ref{tab3} and \ref{tab5} the allowed range for $\alpha_i$ excludes  the $\alpha_1= \alpha_2 =\alpha_3=1$ case.  This result is rigorous and robust in the sense that changing the SUSY breaking scale and the unification scale does not change this conclusion.  Hence, the HSMR parametrization is one of the preferable frameworks to study the RG evolution of Dirac neutrinos now.

%%%%%%%%%%%%%%%%%%%%%%%%%%%%%%%%%%%%%%%%%%%%%%%%%%%%%%%%%%%%%%%%%%%%%%%%%%%%%%%%%%%%%%%%%%%%%%%%%%%%%%%%%%%%%%%%%%%%%%%%%%%%%%%%%%%%%%%%%%%%%%%%%%%%%%%%%%%%%%%%%%%%%%%%%%%%%%%%%%%%%%%%%%%%%%%%%%%%%%%%%%%%%%%%%%%%%%%%%%%%%%%%%%%%%%%%%%%%%%%%%%%%%%%%%%%%%%%

\begin{acknowledgments}
The work of G. A. and M. Z. A. is supported by the Spanish Government and ERDF funds from the EU Commission
[Grants No. FPA2011-23778, FPA2014-53631-C2-1-P and No. CSD2007-00042 (Consolider Project CPAN)].  RS is funded by the Spanish grants FPA2014-58183-P, Multidark CSD2009-00064, SEV-2014-0398 (MINECO) and PROMETEOII/2014/084 (Generalitat Valenciana).
\end{acknowledgments}

%%%%%%%%%%%%%%%%%%%%%%%%%%%%%%%%%%%%%%%%%%%%%%%%%%%%%%%%%%%%%%%%%%%%%%%%%%%%%%%%%%%%%%%%%%%%%%%%%%%%%%%%%%%%%%%%%%%%%%%%%%%%%%%%%%%%%%%%%%%%%%%%%%%%%%%%%%%%%%%%%%%%%%%%%%%%%%%%%%%%%%%%%%%%%%%%%%%%%%%%%%%%%%%%%%%%%%%%%%%%%%%%%%%%%%%%%%%%%%%%%%%%%%%%%%%%%%%

\end{document}